\documentclass{article}

\usepackage{microtype}
\usepackage{graphicx}
\usepackage{subcaption}
\usepackage{booktabs}
\usepackage{hyperref}

 \usepackage[accepted]{icml2026} 
\usepackage{amsmath}
\usepackage{amssymb}
\usepackage{mathtools}
\usepackage{amsthm}
\usepackage[capitalize,noabbrev]{cleveref}
\usepackage{enumitem}
\usepackage{csquotes}

\theoremstyle{plain}

\theoremstyle{definition}

\theoremstyle{remark}

\usepackage[textsize=tiny]{todonotes}

\icmltitlerunning{International Agreements to Limit Frontier AI}

\usepackage{xurl}
\usepackage{etoolbox}
\appto{\bibsetup}{\sloppy}

\begin{document}

\twocolumn[
\icmltitle{International Agreements to Limit Frontier AI: Objectives and Exit}

\begin{icmlauthorlist}
\icmlauthor{Lennart Finke}{eth,harvard,mats}
\icmlaffiliation{eth}{ETH Zürich, Zurich, Switzerland}
\icmlaffiliation{harvard}{Harvard University, Cambridge, United States}
\icmlaffiliation{mats}{MATS Research, Berkeley, United States}
\icmlcorrespondingauthor{Lennart Finke}{lfinke@ethz.ch}
\end{icmlauthorlist}

\icmlkeywords{AI governance, international agreements, frontier AI, treaty design, AI safety}

\vskip 0.3in
]

\printAffiliationsAndNotice{}

\begin{abstract}
An international agreement to limit AI development could be crucial to mitigate risks from AI. However, it remains unclear which conditions should determine when the limiting measures are relaxed. We survey existing international agreements, outline what properties appropriate conditions should satisfy, list possible conditions, and finally give a recommendation in an example scenario. We recommend a fixed time period after which a new organization established at the start of the period specifies conditions that address when AI development can be safely conducted, with a possibility of withdrawal in extraordinary circumstances. We hope to illustrate the considerations that would likely go into an international agreement to limit AI.
\end{abstract}

\section{Introduction}

Both large risks and large gains are expected to stem from the development of increasingly powerful AI. To lower the risks from future AI systems, some have put forward proposals to limit the development of frontier AI, such as \citet{scher2025internationalagreement,miotti2023computecap,alaga2023coordinatedpausing,scholefield2025conditionaltreaty}. Public support for limiting AI development appears plausible in countries where data is available, such as the United States and the United Kingdom \citep{elsey2023uspublicopinion,perrigo2025britsbanai,brysongillette2025opinionsadvancedai}, though data in most countries is sparse. However, broad restrictions of AI development would forgo potential benefits of AI development, which leads most proposals to limit AI development to envision an end of the agreement, as in \citet{aschenbrenner2024existentialrisk}. Here we investigate how the signatories that enter into the agreement will likely think about the path towards the end of the agreement.

The mechanism to weaken or end the agreement has received little attention, as initial works have naturally focused on the restrictions themselves. Nonetheless, it would be a vital consideration for the parties entering the agreement, and some see this lack of consensus as a barrier to a potential agreement.

Since the design of an end phase is tied to the context of the agreement, we must specify which methods of limiting AI development we assume. For simplicity, unless stated otherwise, we assume a similar method to \citet{scher2025internationalagreement}, namely an upper limit on the amount of computational power deployed on training an AI. This method will not pause AI development even if the upper limit on computational power is lower than what is commonly used, because other factors like training data quality and algorithmic improvements will lead to further improvements overall.

We further assume that the agreement is signed by the NATO member states and China, but that there may remain non-signatory states with much less but non-negligible capability to develop frontier AI. Note that we understand the important risks from AI development to put all of humanity at risk, not only the states that conduct AI development, and that we therefore expect signatories to exert pressure on other states to sign as well. A likely scenario would therefore be that all states that could house frontier AI development become signatories. We still assume a more pessimistic scenario to show that the agreement discussed is robust to it.

Apart from the text of the agreement itself, it is important to consider the context in which its signatories enter into it. One crucial aspect is the expectation of countermeasures should a signatory violate the agreement. The existing literature on AI development agreements envisions either a scenario close to all other international agreements, where violations might lead to sanctions or escalation to the United Nations Security Council, or a more favorable scenario where signatories collectively hold the keys to each other's AI infrastructure and can unilaterally halt AI development by withholding their key \citep{aarne2025hardwareguarantees,scher2025verification}. As above, while we hope for the implementation of infrastructure that enables enforcement of an agreement, we assume the more pessimistic case throughout, in which signatories do not have the capacity to unilaterally halt AI development. The great powers that enter into the agreement would then be bound by only minor material factors, and instead the main reason to comply is the reputational cost of a violation. Reputation is especially valuable if there are continued gains from trade between states with and without advanced AI, which is uncertain.

The setting, an upper limit on computational power used for training an AI, offers a straightforward model of relaxing the restrictions imposed by the agreement: Once a certain condition is met, the limit is increased. Raising the limit to a sufficiently large amount of computational power would be equivalent to terminating the agreement entirely. We envision a series of gradual increases. Though the agreement will not be exited completely, we call a condition to increase the limit a \emph{relaxation condition}, because such an increase would mark the exit of one phase of the agreement to the next. In contrast, we call a condition that definitively ends the agreement a \emph{termination condition}. We think of termination conditions as a subset of relaxation conditions.

While we will treat relaxation conditions as relaxing all of the agreement, or the most important restriction of the agreement, we also draw attention to the existing notion of treaty reservation provisions \citep{international2011guide}, which cover any kind of possibility of modification of the treaty after its entering into force. Relaxation conditions can be understood as a subcategory of reservation provisions.

The decision to exit from an agreement can occur in one of two scenarios. First, the purpose of the agreement has been achieved, or second, a signatory wishes to exit even though the original purpose has not been achieved. The agreement should be set up in such a way that the first scenario becomes more likely, and the second less likely.

In this work, then, we survey existing proposals and related international agreements for termination conditions and their implementations, lay out criteria for appropriate relaxation conditions, put forward candidates for conditions, and finally grade the conditions against the criteria.

\section{Termination Conditions in Comparable Agreements}

Previous international agreements have attempted to take countermeasures against other large risks stemming from coordination failures, namely risks from nuclear weapons, biological weapons, and environmental catastrophe. The risk profile of AI development combines aspects of each: A substantial probability of human extinction given unrestricted continued development \citep{grace2025thousands, cais2023statement}, as is the case for nuclear and biological weapons, and possible forgone short-term economic benefits if restrictions are in place \citep{sytsma2025quantifying}, as is the case for environmental protection regulation.

In \Cref{appendix}, we analyze the termination conditions of the Strategic Arms Reduction Treaty, the Washington Naval Treaty, the Treaty on the Non-Proliferation of Nuclear Weapons, the Treaty on the Prohibition of Nuclear Weapons, the Anti-Ballistic Missile Treaty, the Biological Weapons Convention, the Paris Agreement, and the Vienna Convention for the Protection of the Ozone Layer and Montreal Protocol.

None of the surveyed treaties specify particular exit criteria under which the treaty is voided entirely. Instead, they mainly specify how the case is to be treated where a state wishes to exit the treaty without the goal having been satisfied.

\section{Desiderata for Relaxation Conditions}

Systematizing the selection of relaxation conditions, we propose a set of desiderata that a condition might satisfy. We intend, but do not suppose, to be exhaustive in these criteria.

\subsection{Accuracy}

An ideal relaxation condition should capture whether the risks of developing advanced AI have been adequately addressed in the time since the agreement came into effect. This might be difficult to determine, as it requires the signatories to not only agree on the goal of the agreement (to limit AI development), but also on the underlying reason, which is currently viewed as multifaceted. Specific risks that the signatories might want to reduce include risks from autonomous AI entities up to including human extinction, military AI and failure of mutual deterrence, catastrophic misuse like engineered pandemics, or simply economic issues such as rising unemployment.

Given that many non-state entities have an economic stake in the development of frontier AI, it would be expected that those entities would direct substantial efforts toward satisfying the relaxation condition irrespective of the intent behind it. Conversely, an actor may dedicate resources to keep the agreement in place, even if safe and beneficial AI development becomes possible. A relaxation condition ideally captures the true level of risk even under such adversarial conditions.

\subsection{Verifiability}

It should be unambiguous when the relaxation condition has been fulfilled. This property generally trades off against accuracy, because progress in research or societal resilience are difficult to determine unambiguously.

Even when a condition is unambiguous in its definition, it may still be difficult to produce adequate evidence trusted by all signatories. With regard to the verification of the restrictions themselves, see \citet{scher2025verification} and \citet{baker2023verificationlessons}, though the mechanisms necessary to verify relaxation conditions might not overlap much.

\subsection{Preservation of Sovereignty}

While international agreements generally are not understood to supplant national sovereignty, the ``complete autonomy and self-determination over its internal and external affairs'' \citep{ebsco2026sovereignstate}, in practice international agreements constrain how states can exercise their sovereignty. Especially in the context of AI development, where a signatory's assessment of AI risks and benefits likely changes more quickly than in other policy domains, signatories likely prefer agreements that leave open a plausible path to exit the agreement in case their assessment changes.

If not enough sovereignty is preserved within the agreement, it becomes more likely that a signatory violates the agreement. This threshold is low in our assumed scenario without chip infrastructure to halt other signatory's development.

For a full discussion of withdrawal conditions as an incentive to participate in a treaty, see \citet{koremenos2010exit}.

\subsection{Precedent}

Signatory states may place more confidence in a condition with successful precedent in previous international agreements, and may disfavor previously unsuccessful conditions.

\subsection{Prioritization of Desiderata}

Insofar as the above desiderata trade off against each other, it is important to determine whether certain desiderata carry more weight than others. This problem is adequately addressed by modeling the objectives of the signatories. One approach to doing so might be to treat the most important nations as rational actors, assume quantifiable beliefs about AI risk and potential benefits, the likelihood of developing a certain level of AI capabilities before the other actors, and so on, and then assign utility to each desideratum. For the sake of simplicity, we are content with qualitatively laying out the considerations that would go into such a model.

The incentives of signatories at the time of entering on the agreement may depend sensitively on the gap in AI capabilities between the leading and next closest state. We assume that the existing trends continue, with the United States as the leading state producing AI models that are around 7 months ahead of the models by the next closest state, China \citep{epoch2026usvschinaeci}. A larger gap allows for more accuracy at the expense of precedent and verifiability, assuming the leading state is most interested in addressing risks from AI.

The signatory states may have stronger or weaker beliefs in the risks of AI development. In the case of nuclear weapons, many actors signing international treaties can be assumed to have a stronger belief that nuclear war would pose an unacceptable risk to humanity or their citizens in the absence of international agreements. From an address to the UN General Assembly by then U.S. president John F.~Kennedy: ``Mankind must put an end to war---or war will put an end to mankind'' \citep{kennedy1961unga}. Separately, Kennedy is quoted as estimating at least a one-in-three chance that the Cuban Missile Crisis would have led to nuclear war \citep{jfklibrary2002cubanmissile}. As a result of this strong belief, signatories to treaties like the NPT have acceded to verification mechanisms that might otherwise be perceived as expensive or in conflict with national sovereignty, such as inspections by the International Atomic Energy Agency. This level of control by an international body as a result of a voluntary agreement had been unprecedented.

We assume that, given a stronger belief in large-scale AI risk, the desiderata preservation of sovereignty, verifiability, and precedent are considered less important.

Signatories may further have a higher or lower degree of belief in the benefits of AI development. If high, accuracy carries more weight, as the safe relaxation of the agreement is expected to confer larger benefits.

We assume that the signatories treat risk from AI development as comparable to risk from nuclear warfare in the 1960s. This assumption seems justified as a plurality of leading figures in AI from both the United States and China have already signed a statement to this effect \citep{cais2023statement}, and that conditional on states being willing to negotiate an agreement to limit AI development, they would be likely to share this conviction.
We further assume that the signatories have similar beliefs about the economic benefits of unrestricted AI development, which we take to be a substantial, sustained increase in GDP growth rates.

Based on the above considerations, the desiderata might be weighed as: $\text{Accuracy} > \text{Preservation of Sovereignty} > \text{Verifiability} > \text{Precedent}$.

\section{Types of Relaxation Conditions}

\subsection{Absence of a Relaxation Condition}

Some authors have proposed that it is advisable to never develop certain levels of AI capability \citep{aguirre2023keepthefuturehuman}. This stance implies that an agreement should contain no relaxation and in particular no termination condition.

The option to terminate is generally present in international agreements similar to the one we consider here. Nonetheless, agreements without termination conditions are possible. An example of an agreement without termination conditions is the UN Charter \citep{un1945charter}. The Vienna Convention on the Law of Treaties (not to be confused with the treaty concerning ozone depletion) \citep{vienna1969lawoftreaties}, Article 54, reads:

\begin{displayquote}
``The termination of a treaty or the withdrawal of a party may take place:
(a) in conformity with the provisions of the treaty; or
(b) at any time by consent of all the parties after consultation with the other contracting States.''
\end{displayquote}

Article 56, paragraph 1, reads:

\begin{displayquote}
``A treaty which contains no provision regarding its termination and which does not provide for denunciation or withdrawal is not subject to denunciation or withdrawal unless:
(a) it is established that the parties intended to admit the possibility of denunciation or withdrawal; or
(b) a right of denunciation or withdrawal may be implied by the nature of the treaty.''
\end{displayquote}

And Article 62, paragraph 1, reads:

\begin{displayquote}
    A fundamental change of circumstances which has occurred with regard to those existing at the time of the conclusion of a treaty, and which was not foreseen by the parties, may not be invoked as a ground for terminating or withdrawing from the treaty unless: (a) the existence of those circumstances constituted an essential basis of the consent of the parties to be bound by the treaty; and (b) the effect of the change is radically to transform the extent of obligations still to be performed under the treaty.
\end{displayquote}

While it is naturally difficult to anticipate what the most likely scenarios to satisfy Article 62, paragraph 1(a) and (b) are, we believe one likely scenario would be a non-signatory state gaining a realistic chance to develop more advanced AI than the signatory states. 

For the absence of a condition, accuracy is medium. Risk from AI development is fully addressed, at least if consent of all parties  adequately reflects that a reduction in risk from AI development has indeed occurred during the period of the agreement. On the other hand, the agreement may continue even if the risks of AI are addressed. A more formal codification of the conditions under which the risk would be reduced could provide yet higher accuracy. Precedent and preservation of sovereignty are low, while verifiability is by construction high.

\subsection{Fixed Time Limit}

A fixed time-limit expiry ending with a conference where the signatories vote on an extension may be a simple solution. Its main advantages are trivially high verifiability, feasibility, and good precedent. Furthermore, it could be used as a stopgap measure while more nuanced relaxation conditions are developed.

Assuming AI developers are confident that the agreement will not be renewed, they could further use the time before expiry to prepare for increased development efforts, compromising accuracy. The limiting measures of the agreement might consequently have to be amended to cover this effect, increasing the difficulty of enforcing the agreement. Further, the end of the agreement might present a coordination point for entities to influence the states' renewal decision. 

\subsection{Individual Withdrawal / Extraordinary Events}

A condition allowing signatory states to withdraw from the agreement is included in many international treaties to preserve the signatories' sovereignty. Such a withdrawal can be made conditional on the withdrawing signatory disclosing some extraordinary events that caused it to withdraw, as is done in the NPT and ABM treaties. While such a condition has been included in existing agreements, precedent cannot unequivocally be said to be positive, as there are prominent cases of signatories disagreeing about whether another signatory's withdrawal is legitimate, creating tension. For example, the United Nations does not recognize North Korea's withdrawal from the NPT as legitimate. How a withdrawal deemed inconsistent with the treaty by other signatories would be handled is beyond the scope of this document.

Withdrawal clauses usually contain a notice period, and so a natural question is how long this should be for the agreement in question. We imagine the situation under which such a clause would be triggered. While this is challenging, we find the following scenario to be representative: a non-signatory state that was previously significantly behind acquires the chips necessary to conduct a training run at the limit specified in the agreement, thus putting it on roughly equal footing with the leading signatories. It is currently starting to build the facilities. The signatories should at least have enough time to stay ahead of the non-signatory in this scenario. Assuming the computation limit is roughly the contribution of a single U.S. data center for a current frontier training run, the fastest planned build time for a gigawatt-scale frontier data center is 12 months, and the fastest buildout of a frontier data center facility to date was completed in around 4 months \citep{epoch2026frontierdatacenters}. Based on this, we recommend a notice period of 4 months. Note that this assumes that the signatories do not have a reliable way to prevent or predict compute capacity increases of non-signatory states, which might well be the case if large-scale compute monitoring is implemented. It also assumes that frontier data center buildout times will not change; in particular, buildout times could become shorter if it becomes easier to build a data center at the smallest scale that the agreement restricts.
A notice period of 4 months is the 28th percentile of notice periods among 86 international agreements analyzed in \citet{koremenos2010exit}.

\subsection{Progress in Preparedness}

We summarize the totality of approaches to mitigate risks from AI in the term \emph{preparedness}. Preparedness includes AI safety research, such as alignment and interpretability, as well as societal resilience to effects of AI development, such as increased cyberattack capabilities and economic upheaval.

As progress in preparedness continues, the risk from AI development might be sufficiently lowered to continue development. How to best determine this is an open problem, however.

\subsubsection{Expert Consensus}

During the time of the agreement, the scientific community may come to the conclusion that the risk from AI development has been sufficiently lowered, serving as an indication that the agreement may be safely terminated. As one example of several ways to reduce risk, \citet{scher2025internationalagreement} propose an ``established consensus about the efficacy of ASI-alignment methods'' as a condition. The signatories may appoint a council of experts in the field of AI safety and condition relaxation on the council's approval after regular meetings.

What remains to specify is what, exactly, the signatories would task the expert council to answer. One might task it with an end-to-end prediction of the likelihood of catastrophic harm from AI (say, the likelihood of human extinction) from unrestricted AI development over the coming years (say, 25 years), and terminate the agreement if this likelihood is substantially lower with the agreement in place than with the agreement terminated. However, there are several reasons why a signatory might disagree with the expert council on AI risk once it is established. First, each signatory and the council might have different views on the contributions of different risk factors, for instance weighing the prospect of large-scale unemployment against more catastrophic harms like risk from rogue AI systems. Second, the signatories might also collectively be less or more conservative in their assessment of AI risk than the council. These possibilities would be perceived as a limit on the sovereignty of signatories, rendering them less likely to enter into the agreement.

Alternatively, the council might be tasked with judging solutions to more specific questions that the signatories believe to be central to the safe development of AI. One could imagine drawing from the field of pure mathematics, specifying a list of problems akin to the Millennium Prize Problems that, once solved, provide a foundation for safe AI development. One could imagine, for instance, that the expert council asks for progress in AI with provable guarantees.

We present \citet{tegmark2023provablysafe} as a representative effort to characterize what constitutes sufficient progress in AI with provable guarantees.  They outline an agenda to develop AI systems that provably satisfy safety requirements by extracting algorithms from neural network weights using interpretability techniques, then verifying the adherence of the extracted algorithm to requirements by formulating the requirements as proof-carrying code. Such proofs could conceivably rule out that an AI produces harmful output, or mathematically bound the probability with which an AI produces harmful output. The agenda also identifies interventions at several other layers of AI development to prevent execution of unsafe AI systems. A relaxation condition based on these proposals might require the development of all of the above for a class of AI systems that can be expected to yield a significant fraction of economic benefit as compared to what unrestricted AI systems would yield. A more comprehensive overview of approaches to provably safe AI is given in \citet{dalrymple2024guaranteedsafeai}.

Given that many actors in AI development today explicitly or implicitly treat provably safe AI systems as intractable, we present two other representative types of conditions that have recently emerged out of practice.

With regard to risk of catastrophic AI misuse, \citet{zou2025constitutionalclassifiers} give a practical definition for robustness against adversarial attacks. They let well-incentivized testers spend a fixed amount of time trying to find an input that makes their system output harmful content. While they made significant advancements, and could therefore report that no tester elicited harmful output on all 10 test cases, the testers did elicit harmful output in many cases. Further, testing was done for a cumulative 3,000 hours across all testers, which is still below what the model is exposed to in deployment. When exposing the system to testers through an openly accessible challenge, multiple testers found strategies that produced harmful input on all 10 test cases within the span of 5 days. A relaxation condition based on this line of research might ask for development of a system that performs above a given capability threshold, while producing no harmful output in a test like the above.

Largely aiming to address risks from rogue AI, \citet{olah2023interpretabilitydreams} and \citet{elhage2022toymodels} outline an agenda to simplify the problem of fully characterizing likely outputs from an AI. They claim that this may be possible by enumerating some distinct concepts, or ``features'', that make up the computation that leads to the output. If those features are exhaustively enumerated and characterized by when they arise, and in turn satisfactorily explain the output of the system, this would mark significant progress in mitigating many types of AI risk. While there is no agreed-upon precise operationalization of the above argument that could readily serve as a condition, most researchers in interpretability today work within this paradigm \citep{sharkey2025openproblems}. A relaxation condition based on this line of research might ask for an AI model that performs above a given capability threshold, with outputs explainable by an exhaustive set of intelligible features.

Progress in preparedness for the development of advanced AI encompasses more than AI research alone. For example, sufficiently advanced AI could cause rapid job displacement \citep{occhipinti2025socioeconomictippingpoint,korinek2024scenarios}, or disruption of international relations \citep{pavel2025nations}. We therefore believe it would be beneficial for states to develop contingency plans for advanced AI as it relates to their internal affairs, as well as publish plans on how they will act internationally once advanced AI is developed. A relaxation condition based on this might necessitate submission of a report in a specified format by each signatory.

We do not recommend specific conditions here out of those sketched above. None of the conditions presented here are to be understood as completely sufficient conditions to nullify risks from AI development in isolation.

\subsection{Deference to a Later Specification}
\label{sec:later-specification}

Given the difficulty of accurately specifying an adequate level of preparedness, one can imagine a two-stage agreement, the first stage of which terminates when such a specification has been produced.

Since the creation of such a specification is sufficiently difficult to warrant coordinated research efforts, this could be understood analogously to the Vienna Convention, in which the signatories were not sufficiently certain to commit to specific measures, and instead decided to commit to researching appropriate measures instead. The crucial difference between AI development and ozone depletion, besides the much larger scale of risk from AI development, is the high speed and exponential nature of AI development compared to the low speed and approximately linear nature of ozone depletion. Therefore, for the agreement we treat here, we could require that restrictions are already in place for the first stage of the agreement, whereas the Vienna Convention did not impose any binding restrictions to lower ozone-depleting emissions. Instead of deferring the restrictions themselves to the second stage, we defer the relaxation conditions.

Observing a period in which the restrictions of the agreement are in place might enable a more accurate specification of a relaxation condition, as this allows gathering more data on how preparedness research is impacted by the restrictions.

\subsection{Absence of AI-Related Incidents}

While the development of frontier AI systems would be restricted under the agreement, it is likely that the deployment of AI systems at least as powerful as those available today remains unrestricted. Therefore, incidents of harm from AI can provide an indication that relaxation would not be appropriate, and conversely, absence of such incidents provides weak evidence that relaxation is more appropriate than previously thought. Relatedly, \citet{scher2025internationalagreement} propose ``strong misuse and proliferation controls'' as a relaxation condition, which might be empirically evaluated by their efficacy in preventing incidents with current technology. Damages from AI-related incidents can be measured with existing liability infrastructure, though it might be unclear what incident is attributable to an AI system as opposed to its user, owner, or manufacturer \citep{smith2024liability}.

On the other hand, this approach is vulnerable to manipulation by an entity wishing to prevent relaxation of the agreement via purposefully instigating AI-related incidents. Also, and most importantly, it is far from clear that the absence of AI-related incidents under the agreement restrictions provides sufficient evidence that development without the restrictions would continue to be safe.

One example of an entity already tracking AI-related incidents is the AI Risk Repository, which counted 370 incidents in 2025, of which 211 were attributed to malicious actors \citep{slattery2024airiskrepository}.

\subsection{Creation of a Global AI Development Project}

Some of the risks stemming from AI development are mediated by the fact that current frontier AI development efforts are undertaken by several distinct companies, which treat each other as competitors within and beyond state borders. The current situation is adequately described as a race of multiple actors who individually wish, or claim to wish, to slow development and spend significantly more resources on safe development, but cannot, because they would be overtaken by the other actors, losing their upside. A unified AI development project would allow these actors to balance AI development and safety research to a level deemed more appropriate, enable a more democratic AI development process, and allow the signatories to continue to impose restrictions on the unified project as needed. \citet{scher2025internationalagreement} propose the ``creation of a globally-monitored AI development project that proceeds with due caution'' as one of the necessary conditions to terminate the agreement. We assume that actors besides the international project would still be restricted by the original conditions.

The creation of a global AI development project only indirectly addresses risks from AI development, by enabling resources to be adequately allocated to preparedness research. As in \Cref{sec:later-specification}, the project would still need to determine under what conditions additional development restrictions would need to be initiated and terminated, albeit under significantly more favorable conditions.

One downside of this approach is that a global AI development project would require a perhaps unprecedented level of international coordination of state actors and, once successful, could become a powerful entity in comparison to a single signatory state. Therefore, the creation of such an entity would be an unusual policy goal for any one party, and certainly for most states taken together, to pursue, which in turn might mean that the relaxation condition would be unduly unlikely to be satisfied. If the condition includes a requirement that a certain state must be part of the development project for the agreement to terminate, that state would have considerable influence over keeping the agreement in force.

An overview of our judgment for each relaxation condition type on each desideratum is given in \Cref{tab:conditions}.

\section{Future Work}

As current efforts to establish an international agreement are still in very early stages, we are tempted to recommend various lines of future work. We limit ourselves to mentioning only those that we expect to be immediately useful given the unprecedented speed of AI development \citep{kwa2025longsoftwaretasks}.

A consensus on what constitutes sufficient conditions to continue AI development could be reached in the near future. One simple step towards this could be a survey of AI experts in the United States and China on what they believe AI risks are and which methods to address each risk are most promising. While we believe that such a survey would currently yield many contradictory responses and no clear majority, it could also uncover unexpected commonalities within and across state borders. We think that beliefs around AI risk will converge over time as risks become more apparent. Beyond a passive assessment of researchers' stances on AI risk, dialogue between experts would likely accelerate consensus.

In a dialogue during the 2026 World Economic Forum, the CEOs of Anthropic and Google DeepMind agreed that both were in favor of a slower pace of development, but showed concern about the feasibility of an international agreement \citep{miailhe2026slowdownai}. We encourage frontier AI companies to develop their existing necessary conditions for release of a current model generation into sufficient conditions that would be suitable as relaxation conditions in an international agreement. We believe that engagement from AI companies on questions of international coordination would be highly promising in general.

\section{Conclusion}

We give a recommendation for relaxation conditions based on our analysis.

The agreement should end after 5 years, during which time the organization determining satisfaction of the relaxation conditions should be established. This organization proposes the relaxation conditions, and if the parties cannot agree to them after 5 years, they may agree to extend the agreement by another 2 years, or terminate it. The accepted relaxation conditions must address preparedness for the possibility of rogue AI, catastrophic misuse, and catastrophic societal developments. Individual withdrawal based on extraordinary events is possible with 4 months' notice.

We briefly give our reasoning for each element of the agreement. The creation of an organization that specifies relaxation conditions is based on the construction from the Vienna Convention (see \Cref{sec:later-specification}) and might cease to be necessary once a consensus on adequate preparedness emerges among experts. Requiring parties to later agree on the terms, as opposed to a unilateral decision by the newly created organization, prevents conditions that are either too lax or too strict: we expect some signatories to have stronger, and some weaker, beliefs regarding AI risk. In case it is deemed too unlikely that all signatories agree, a possible variation would be to designate a subset of signatories whose agreement is sufficient for the agreement to proceed to the next phase. We include the withdrawal clause largely based on the assumption that a non-signatory state has a realistic chance to develop frontier AI. If such a non-signatory state were not to exist, we would not recommend a withdrawal clause, based on the relative success of the Washington Naval Treaty and controversial uses in other agreements, like the ABM treaty and NPT.

We stress again that the above is based on one specific proposal for restrictions on AI development, a compute limit. Though an unsolved problem, we believe it to also be possible to impose restrictions that do not need relaxation conditions, because they accurately capture the risks of advanced AI without the need for a proxy. For instance, one could imagine international restrictions of AI systems based on their capability; for example an AI system that can autonomously self-replicate, that can autonomously perform a large fraction of cognitive labor of today, or that can autonomously perform cyber attacks at a certain level of capability. This type of restriction would be more similar to frontier safety policies already implemented at companies \citep{metr-2025-common-elements-of-frontier-ai-safety-policies}. Because this approach comes with many challenges of its own, such as the lack of an adversarially robust specification of levels of capability, or disagreement about what capabilities are economically valuable enough to keep unrestricted, we find relaxation conditions in conjunction to more verifiable restrictions to be a promising avenue.

We look forward to continued, increasingly practical progress in international cooperation on AI. We believe that an international agreement as discussed would adequately address the single largest risk that each potential signatory state, and humanity as a whole, faces today. 

\newpage

\section*{Acknowledgements} We greatly appreciate comments by Aaron Scher, David Krueger, Mauricio Baker, Oliver Guest, Matthijs Maas, Thomas Larsen, and Nate Soares, which significantly improved this work. All opinions expressed are the authors', and in no way reflect those of the acknowledged. All errors are the authors'.

\section*{Impact Statement}
We write this work in hope of having a positive impact on the trajectory of AI development. Some potential negative side effects of our proposal, such as forgone short-term economic benefits, are discussed in the main text. We believe that international agreement to limit AI development would be highly beneficial overall.

\section*{AI Usage Statement}
The text in this article was written manually by the authors. Language models were used for spell-checking, some copy-editing, literature search, and formatting, though all resulting edits were applied manually.

\begin{table*}[t]
\caption{Qualitative assessment of relaxation condition types against the desiderata.}
\label{tab:conditions}
\centering
\small
\begin{tabular}{lcccc}
\toprule
\multicolumn{1}{c}{} & \multicolumn{4}{c}{\textbf{Desideratum}} \\
\cmidrule(lr){2-5}
\textbf{Relaxation Condition} & \textbf{Accuracy} & \textbf{Verifiability} & \textbf{Preservation of Sovereignty} & \textbf{Precedent} \\
\midrule
Absence of a Relaxation Condition & Medium & High & Low & Medium \\
Fixed Time Limit & Low & High & Medium & High \\
Extraordinary Events & Low & Low & High & Medium \\
Absence of AI-Related Incidents & Medium & High & Medium & Low \\
Creation of a Global AI Development Project & Medium & High & Low & Low \\
Progress in Preparedness & High & Medium & Medium & Low \\
Deference to a Later Specification & N/A & N/A & High & Medium \\
\bottomrule
\end{tabular}
\end{table*}

\appendix
\section{Comparable Treaties} \label{appendix}
\subsection{Strategic Arms Reduction Treaty}

The Strategic Arms Reduction Treaty (START) \citep{start1991} is perhaps the closest international agreement to what we treat here. In it, the United States and Soviet Union coordinate to reduce their capabilities to launch nuclear weapons. The agreement contains detailed technical requirements and sophisticated mechanisms of verifying compliance. Its Article XVII specifies the termination conditions. It includes a 15-year time limit after which a 5-year extension is to be agreed on, and a right to withdraw from the treaty based on ``extraordinary events, related to the subject matter of this Treaty'' that ``have jeopardized the supreme interests'' of the signatory country, with 6 months’ notice. Before the 15 years ran out, the Soviet Union had collapsed, and a new agreement was reached between the United States and Russia, the Strategic Offensive Reductions Treaty. This was in turn superseded by the Measures for the Further Reduction and Limitation of Strategic Offensive Arms Treaty (New START) \citep{newstart2010}. The latter has identical termination conditions to START, except with 3 months’ notice and a 10-year renewal period and a renewal of ``no more than five years''. One month before expiry, the agreement was renewed by 5 years, though Russia later announced that it was suspending its participation before expiry, alleging non-compliance by the United States \citep{afp2023putinnewstart}. The agreement has now expired.

\subsection{Washington Naval Treaty}

The Washington Naval Treaty \citep{washington1922naval} coordinated the slowing of warship development between the Allies after World War~I. While old, it is relevant to the agreement we treat because it restricts as of then undeveloped technology in a quantitative manner, i.e., it restricts displacement volume of warships, barrel diameters, and similar characteristics. Its Article XXIII treats termination. After a minimum duration of around 14 years, a signatory could give unilateral notice of termination without providing a reason for doing so, subject to a two-year notice period. The treaty would then terminate within two years of receipt of the notice. After several parties violated the agreement either only in spirit or fully, Japan gave notice of termination at the end of the 24-year period.

\subsection{Treaty on the Non-Proliferation of Nuclear Weapons}

The Treaty on the Non-Proliferation of Nuclear Weapons (NPT) coordinates an effort to stop the spread of nuclear weapons among 190 states \citep{npt1968}. Notably, India, Pakistan, and Israel did not sign the treaty. India and Pakistan first demonstrated functional nuclear weapons after NPT went into force, whereas Israel is commonly believed to have developed functional nuclear weapons since before NPT was signed. The treaty's Article X specifies the termination conditions, which are similar to START except for a 25-year initial duration and a 6 months' notice of withdrawal. Notably, the latter clause was exercised by North Korea in 1993, then reversed before the end of the notice period. It withdrew once more in 2003, citing what it perceived as hostile U.S. foreign policy and an IAEA resolution \citep{dprk2003withdrawal,japanmofa2003iaea}. As the withdrawal is not recognized by many other signatories, it is considered by some states, including the United States, to be in violation of the treaty \citep{state2025compliance}. No other signatory state has exercised the article.

\subsection{Treaty on the Prohibition of Nuclear Weapons}

The Treaty on the Prohibition of Nuclear Weapons expands on the provisions of the NPT by coordinating further, and eventually complete, nuclear disarmament of 95 signatory states \citep{tpnw2017}. No state in possession of nuclear weapons has signed the treaty. Its Article 17 specifies the termination conditions. The duration is explicitly declared as unlimited. Withdrawal is possible under similar conditions as in the NPT, except with a longer 12 months' notice. The withdrawal is considered invalid if the withdrawing signatory is party to an armed conflict, and the signatory remains bound by the treaty until the end of that conflict. No signatory state has exercised the article.

\subsection{Anti-Ballistic Missile Treaty}

The Anti-Ballistic Missile (ABM) Treaty was a bilateral agreement between the United States and the Soviet Union to restrict the development of defense systems against intercontinental ballistic missiles \citep{abm1972}. It is similar to the agreement we assume here in that it voluntarily limits the scale of both existing and future technology, forgoing a potential military advantage over the counterparty. Its Article XV specifies the termination clause. While the duration is explicitly specified as unlimited, the signatories reserve a right to withdraw, again based on ``extraordinary events, related to the subject matter of this Treaty'' that ``have jeopardized the supreme interests'' of the signatory country. In 2002, the United States withdrew. Former president George W.~Bush issued a public statement, distinct from the notification to Russia, citing that the agreement ``hinders our government's ability to develop ways to protect our people from future terrorist or rogue-state missile attacks'' \citep{bush2001abmwithdrawal} and a shifting geopolitical situation after the collapse of the Soviet Union.

\subsection{Biological Weapons Convention}

The Convention on the Prohibition of the Development, Production and Stockpiling of Bacteriological (Biological) and Toxin Weapons and on Their Destruction coordinates the non-proliferation of biological weapons between 189 parties \citep{bwc1972,unoda_treaties_database}. Its Article XIII specifies termination conditions. The agreement is explicitly stated as being unlimited in duration. Similarly to the ABM and NPT treaties, parties can withdraw based on ``extraordinary events'', with a 3 months' notice and a statement to the United Nations Security Council. No signatory has exercised this article.

There have been several cases of parties to the treaty alleging violation of the treaty by other parties. The United States has alleged violation of the treaty by Russia and North Korea, for instance in its 2025 Adherence to and Compliance with Arms Control, Nonproliferation, and Disarmament Agreements and Commitments Report \citep{state2025compliance}. Conversely, Russia has alleged violation by the United States, chiefly in its Outcome Report of the Parliamentary Commission on Investigation into the Circumstances Related to Creation of Biological Laboratories by U.S. Specialists on the Territory of Ukraine \citep{russianmfa2023biolabs}. An instance of a party to the treaty admitting to violation is given by former Chairman of the Supreme Soviet of the Russian SFSR Boris Yeltsin, then president of Russia \citep{cshl2026meselson}.

\subsection{Paris Agreement}

The Paris Agreement coordinates the lowering of greenhouse gas emissions among 194 parties \citep{paris2015agreement}. Its only termination condition is Article 28, which concerns withdrawal from the agreement. Withdrawal is stated as possible starting 3 years after the agreement entered into force, with a 1 year notice period. The United States exercised the article shortly after the agreement entered into force, and once the 3 year period was over, withdrew. It was readmitted into the agreement soon after, but exercised the article again four years later. No other signatory state has exercised the article.

However, many parties are widely understood to be in violation of different requirements of the agreement. For instance, the agreement requires submission of nationally determined contribution reports on emission targets and efforts to reach them. Though submission was due in February 2025, by September 2025 only 57 parties out of 194 had submitted such a report \citep{unep2025emissionsgap}.

\subsection{Vienna Convention for the Protection of the Ozone Layer and Montreal Protocol}

The Vienna Convention for the Protection of the Ozone Layer coordinates the lowering of emissions that deplete atmospheric ozone among 198 ratifying states \citep{vienna1985ozone}. Its Article 19 specifies termination conditions, and is structured identically to the Paris Agreement, except for allowing withdrawal only after 4 instead of 3 years since going into force. No ratifying state has exercised the article.

Importantly, the convention largely leaves the implementation of what precisely is to be restricted to later-determined protocols, mainly focusing on commitment to research efforts to determine appropriate restrictions. This led to the creation of the Montreal Protocol two years later, specifying the corresponding binding restrictions.

The Montreal Protocol on Substances That Deplete the Ozone Layer coordinates 197 ratifying states toward the same goal as the Vienna Convention \citep{montreal1987protocol}. Its Article 19 specifies termination conditions, and inherits the conditions from the Vienna Convention's Article 19. No ratifying state has exercised the article.

\bibliography{biblio}
\bibliographystyle{icml2026}

\end{document}